\documentclass[pra,aps,twocolumn,showpacs]{revtex4}
\usepackage{times,epsfig,amssymb,amsfonts,amsmath,graphicx}

\newcommand{\ket}[1]{|#1\rangle}
\newcommand{\bra}[1]{\langle #1|}

\newcommand{\beq}{\begin{equation}}
\newcommand{\eeq}{\end{equation}}
\newcommand{\beqa}{\begin{eqnarray}}
\newcommand{\eeqa}{\end{eqnarray}}

\begin{document}

\title{General monogamy relation of multi-qubit systems in terms of squared R\'{e}nyi-$\alpha$ entanglement}

\author{Wei Song$^{1}$}
\email{weis@hfnu.edu.cn}
\author{Yan-Kui Bai$^{2}$}
\email{ykbai@semi.ac.cn}
\author{Mou Yang$^{3}$}

\author{Ming Yang$^{4}$}

\author{Zhuo-Liang Cao$^{1}$}

\affiliation{$^{1}$ Institute for Quantum Control and Quantum Information, and School
of Electronic and Information Engineering, Hefei Normal University, Hefei 230601, China\\
$^{2}$College of Physics Science and Information Engineering and Hebei
Advance Thin Films Laboratory, Hebei Normal University, Shijiazhuang, Hebei 050024, China\\
$^{3}$ Laboratory of Quantum Engineering and Quantum Materials, School of Physics and
Telecommunication Engineering, South China Normal University, Guangzhou 510006, China\\
$^{4}$School of Physics and Material Science, Anhui University, Hefei 230601, China}

\begin{abstract}
We prove that the squared R\'{e}nyi-$\alpha$ entanglement (SR$\alpha$E), which is the generalization
of entanglement of formation (EOF), obeys a general monogamy inequality in an arbitrary $N$-qubit mixed
state. Furthermore, for a class of R\'{e}nyi-$\alpha$ entanglement, we prove that the monogamy relations
of the SR$\alpha$E have a hierarchical structure when the $N$-qubit system is divided into $k$ parties.
As a byproduct, the analytical relation between the R\'{e}nyi-$\alpha$ entanglement and the squared
concurrence is derived for bipartite $2\otimes d$ systems. Based on the monogamy properties of
SR$\alpha$E, we can construct the corresponding multipartite entanglement indicators which still work
well even when the indicators based on the squared concurrence and EOF lose their efficacy. In addition,
the monogamy property of the $\mu$-th power of R\'{e}nyi-$\alpha$ entanglement is analyzed.
\end{abstract}


\maketitle

\section{Introduction}

Monogamy of entanglement (MoE) \cite{hor09rmp} is an essential feature in many-body quantum systems,
which means that quantum entanglement cannot be shared freely in multipartite systems \cite{ben96pra}.
Coffman, Kundu, and Wootters established the first quantitative characterization of the MoE for the
squared concurrence (SC) \cite{woo98prl} in an arbitrary three-qubit quantum state \cite{ckw00pra}.
Furthermore, Osborne and Verstraete generalized this monogamy relation to the $N$-qubit case
\cite{osb06prl},
\beq\label{q1}
C^2(\rho_{A_1|A_2\dots A_n})-C^2(\rho_{A_1A_2})-\dots-C^2(\rho_{A_1A_n})\geq 0.
\eeq
where $C^2(\rho_{A_1|A_2\dots A_n})$ quantifies bipartite entanglement in the partition
$A_1|A_2\dots A_n$  and $C^2(\rho_{A_1A_i})$ characterizes two-qubit entanglement with $i=2,3,\dots,n$.
The MoE of SC can be
used to characterize the entanglement structure in multipartite quantum systems and detect the existence
of multiqubit entanglement in dynamical procedures \cite{byw07pra,baw08pra,off08pra,Jung09,eos09pra,
raj10pra,Cornelio13,Regula14,kim14pra,byw09pra,agu14prl}. Moreover, there are also many works devoted to
the topic of entanglement monogamy \cite{csyu05pra,Yang06,Chi08,Yu08,Kay09,Saez13,Osterloh15,Eltschka15}
and similar monogamy relations were also established for Gaussian systems
\cite{ade06njp,hir07prl,ade07prl},
squashed entanglement \cite{koa04pra,chr04jmp,yang09ieee}, entanglement negativity \cite{fan07pra,
kim09pra,he15pra,choi15pra,yluo15ap} and R\'{e}nyi-$\alpha$ entanglement \cite{kim10jpa,cor10pra}.

A genuine three-qubit entanglement measure named ``three-tangle" was obtained from the MoE of SC in
three-qubit pure states \cite{ckw00pra}. However, there exists a kind of three-qubit mixed states which
is entangled but without two-qubit concurrence and three-tangle \cite{loh06prl}, and the similar case
also exists in $N$-qubit systems \cite{byw08pra}. Recently, it was indicated that the squared
entanglement of formation (SEF) \cite{woo98prl} obeys the monogamy relation in multiqubit systems
\cite{bai13pra,oli14pra,bxw14prl,zhu14pra,bxw14pra,gao15sr}. In particular, it was proved analytically that
the SEF is monogamous in an arbitrary $N$-qubit mixed state \cite{bxw14prl},
\beq\label{q2}
E_f^2(\rho_{A_1|A_2\dots A_n})-E_f^2(\rho_{A_1A_2})-\dots-E_f^2(\rho_{A_1A_n})\geq 0,
\eeq
which overcomes the flaw of the MoE of SC and can be utilized to detect all multiqubit entanglement.
R\'{e}nyi-$\alpha$ entanglement (R$\alpha$E) \cite{hor96pla} is also well-defined entanglement measure
which is the generalization of entanglement of formation (EOF) and has the merits for characterizing
quantum phases with differing computational power \cite{jcui12nc}, ground state properties in many-body
systems \cite{fran14prx}, and topologically ordered states \cite{fla09prl,hal13prl,ham13prl,jcui13pra}.
Therefore, it is natural to study the MoE of the R$\alpha$E and its applications in multipartite
entanglement detection. Kim and Sanders proved that the R$\alpha$E with the order $\alpha\geq 2$ obeys
a monogamy inequality in $N$-qubit systems \cite{kim10jpa}, but this monogamy relation does not cover
the case of EOF which corresponds to the R$\alpha$E with the order $\alpha=1$. Whether or not there
exists a general monogamy relation via the R$\alpha$E is yet to be resolved.

In this paper, we analyze the properties of the squared R\'{e}nyi-$\alpha$ entanglement (SR$\alpha$E)
and prove that the SR$\alpha$E with the order $\alpha\geq (\sqrt{7}-1)/2\simeq 0.823$ obeys a general
monogamy relation in an arbitrary $N$-qubit mixed state. This result provides a broad class of new
monogamy inequalities including the monogamy relation of the SEF in Eq. (2) as a special case.
Furthermore, it is proved that the monogamy relations of SR$\alpha$E have a hierarchical structure when
the $N$ qubit systems is divided into $k$ parties. As a byproduct, we give an analytical expression of
the R$\alpha$E as a function of SC in $2\otimes d$ systems. The monogamy relations of the SR$\alpha$E
can be utilized to detect the multipartite entanglement and the SR$\alpha$E-based indicators we
construct can work well even when the corresponding ones based on the SC and SEF lose their efficacy.
Finally, we analyze the monogamy property of the $\mu$-th power of R\'{e}nyi-$\alpha$ entanglement.

\section{Monogamy inequalities for SR$\alpha$E in $N$-qubit systems}

For a bipartite pure state $\left| \psi  \right\rangle _{AB}$, the R$\alpha$E is defined as
\cite{hor96pla}
\beq\label{q3}
E_{\alpha}(\left| \psi  \right\rangle _{AB}):= S_\alpha(\rho_A) = \frac{1}{1-\alpha}\log _2(\mbox{tr}\rho _A^\alpha)
\eeq
where the R\'{e}nyi-$\alpha$ entropy is $S_\alpha(\rho_A)=[\log _2(\sum_i\lambda_i^{\alpha})]/(1-\alpha)$ with
$\alpha$ being a nonnegative real number and $\lambda_i$ being the eigenvalue of reduced density matrix
$\rho_A$. The R\'{e}nyi-$\alpha$ entropy $S_\alpha \left( \rho  \right)$ converges to the von Neumann
entropy when the order $\alpha$ tends to 1. For a bipartite mixed state $\rho _{AB}$, the R$\alpha$E
is defined via the convex-roof extension
\beqa\label{q4}
E_\alpha(\rho _{AB})=\mbox{min} \sum_i p_i E_\alpha(\ket{\psi _i }_{AB})
\eeqa
where the minimum is taken over all possible pure state decompositions of ${\rho_{AB}=\sum\limits_i{p_i
\left| {\psi _i } \right\rangle _{AB} \left\langle {\psi _i } \right|} }$. In particular, for a
two-qubit mixed state, the R$\alpha$E with $\alpha\geq 1$ has an analytical formula which is expressed
as a function of the SC \cite{kim10jpa}
\beqa\label{q5}
E_\alpha \left( {\rho _{AB} } \right) = f_\alpha \left[ {C^2 \left( {\rho _{AB} } \right)} \right]
\eeqa
where the function $f_\alpha \left( x \right)$ has the form
\beq\label{q6}
f_\alpha \! \left( x \right)\!= \!\frac{1}{{1 - \alpha }}\!\log _2 \!\left[ {\left( {\frac{{1 \!-\!
\sqrt {1 - x} }}{2}} \right)^\alpha  \!\!\!\! +\! \left( {\frac{{1 \!+\! \sqrt {1 - x} }}{2}}
\right)^\alpha  } \right].
\eeq
Recently, Wang \emph{et al} further proved that the formula in Eq. (5) holds for the order
$\alpha\geq(\sqrt{7}-1)/2\simeq 0.823$ \cite{yxwang15arx}.

Before presenting the main results of this paper, we first give three lemmas as follows.

\emph{Lemma 1}. The squared R\'{e}nyi-$\alpha$ entanglement $E_\alpha^2 \left( {C^2 } \right)$
with $\alpha\geq (\sqrt{7}-1)/2$ in two-qubit mixed states varies monotonically as a function of
the squared concurrence ${C^2}$.

\emph{Proof}: This lemma holds if the first-order derivative $\partial E_\alpha^2/\partial x > 0$ with
$x = C^2$. After a direct calculation, we have
\beqa\label{q7}
\frac{{\partial E_\alpha ^2 }}{{\partial x}} = \frac{{\alpha\left({B^{\alpha - 1}- A^{\alpha-1}}
\right)\ln \left[{2^{-\alpha} \left( {A^\alpha + B^\alpha} \right)} \right]}}{{\left({1-\alpha }
\right)^2 \left({A^\alpha+B^\alpha}\right)\sqrt {1-x} \left({\ln 2} \right)^2}}
\eeqa
which is always nonnegative for $0 \le x \le 1$ and $\alpha  \ge 0$ with the parameters
$A=1+\sqrt{1-x}$ and $B=1-\sqrt{1-x}$, and the equality holds only at the boundary of $x$. Thus we
obtain that ${E_{\alpha}^2 }$ is monotonically increasing as a function of the squared concurrence,
which completes the proof.

\emph{Lemma 2}. The squared R\'{e}nyi-$\alpha$ entanglement $E_\alpha^2 \left( {C^2 } \right)$ with
$\alpha\geq (\sqrt{7}-1)/2$ is convex as a function of the squared concurrence ${C^2 }$.

\emph{Lemma 3}. The R\'{e}nyi-$\alpha$ entanglement $E_\alpha \left( {C^2 } \right)$ with
$\alpha\in[(\sqrt7-1)/2,(\sqrt{13}-1)/2]$ is monotonic and concave as a function of the squared
concurrence ${C^2 }$.

The proofs for lemma 2 and lemma 3 can be seen from Appendices A and B.

Now, we give the main results of this paper.

\emph{Theorem 1}. For an arbitrary $N$-qubit mixed state $\rho_{A_1A_2\dots A_n}$, the squared R\'{e}nyi-$\alpha$ entanglement satisfies the monogamy relation
\beq\label{q8}
E_\alpha^2(\rho_{A_1|A_2\dots A_n})- E_\alpha^2(\rho_{A_1A_2})-\cdots -E_\alpha^2(\rho_{A_1A_n})\geq 0,
\eeq
where $E^2_\alpha(\rho_{A_1|A_2\dots A_n})$ quantifies the entanglement in the partition
$A_1|A_2\dots A_n$ and $E_\alpha^2(\rho_{A_1A_i})$ quantifies the one in two-qubit subsystem $A_1A_i$
with the order $\alpha\ge (\sqrt 7  - 1)/2$.

\emph{Proof}. We first consider the monogamy relation in an $N$-qubit pure state
$\ket{\psi}_{A_1A_2\dots A_n}$. The entanglement $E_\alpha(\ket{\psi}_{A_1|A_2\dots A_n})$ can be
evaluated using Eq.(\ref{q5}) since the subsystem $A_2\dots A_n$ can be regarded as a logic qubit. Thus,
we can obtain
\beqa\label{q9}
E_\alpha^2(\ket{\psi}_{A_1|A_2\dots A_n})&=&E_\alpha^2[C^2_{A_1|A_2\dots A_n}(\ket{\psi})]\nonumber\\
&\geq& E_\alpha^2\left(\sum_{i=2}^{n} C^2_{A_1A_i}\right)\nonumber\\
&\geq& \sum_{i=2}^n E_{\alpha}^2(\rho_{A_1A_i}),
\eeqa
where in the first inequality we have used the monogamy relation of squared concurrence
$C^2_{A_1|A_2\dots A_n}\geq \sum_{i=2}^n C^2_{A_1A_i}$ \cite{ckw00pra,osb06prl} and the monotonically
increasing property of $E_\alpha^2(C^2)$ (lemma 1), and in the second inequality we have further used
the convex property of $E_\alpha^2(C^2)$ (lemma 2).

Next, we analyze the monogamy relation in an $N$-qubit mixed state $\rho_{A_1A_2\dots A_n}$. In this
case, the formula of R\'{e}nyi-$\alpha$ entanglement in Eq.(\ref{q5}) cannot be applied to
$E_\alpha(\rho_{A_1|A_2\dots A_n})$ since the subsystem $A_2\dots A_n$ is not a logic qubit in general.
But we can still use the formula in Eq.(\ref{q4}) which comes from the convex roof extension of the pure
state entanglement. Therefore, we have
\beq\label{q10}
E_\alpha(\rho_{A_1|A_2\dots A_n})=\mbox{min}\sum p_i E_\alpha(\ket{\psi_i}_{A_1|A_2\dots A_n}),
\eeq
where the minimum is taken over all possible pure state decompositions $\{p_i, \ket{\psi_i}\}$ of
the mixed state $\rho_{A_1A_2\dots A_n}$. Assuming that the optimal decomposition for Eq.(\ref{q10}) is
$\rho_{A_1A_2\dots A_n}=\sum_{j=1}^{m}p_j \ket{\psi_j}_{A_1A_2\dots A_n}\bra{\psi_j}$, we have
\beqa\label{q11}
E_\alpha^2(\rho_{A_1|A_2\dots A_n})&=&[\sum_j p_j E_\alpha(\ket{\psi_j}_{A_1|A_2\dots A_n})]^2\nonumber\\
&=&\{\sum_j p_j E_\alpha[C_{A_1|A_2\dots A_n}(\ket{\psi_j})]\}^2\nonumber\\
&\geq& \{E_\alpha[\sum_j p_j C_{A_1|A_2\dots A_n}(\ket{\psi_j})]\}^2\nonumber\\
&\geq& \{E_\alpha[C_{A_1|A_2\dots A_n}(\rho)]\}^2\nonumber\\
&=&E_\alpha^2[C^2_{A_1|A_2\dots A_n}(\rho)]\nonumber\\
&\geq& E^2_\alpha[\sum_{i=2}^n C^2(\rho_{A_1A_i})]\nonumber\\
&\geq& \sum_{i=2}^{n} E_\alpha^2[C^2(\rho_{A_1A_i})]\nonumber\\
&=&\sum_{i=2}^{n} E_\alpha^2(\rho_{A_1A_i}),
\eeqa
where we have used in the second equality the pure state formula of the R$\alpha$E and taken the
$E_\alpha(C)$ as a function of the concurrence $C$ for $\alpha \ge (\sqrt {7} - 1)/2$; in the
third inequality we have used the monotonically increasing and convex properties of $E_\alpha(C)$ as a
function of the concurrence \cite{kim10jpa}; in the forth inequality we have used the convex property
of concurrence for mixed states; and in the sixth and seventh inequalities we use the monotonically
increasing and convex properties of $E^2_\alpha(C^2)$ as a function of the squared concurrence
(lemmas 1 and 2). Thus we have completed the proof of theorem 1.

\emph{Theorem 2}. For a bipartite $2 \otimes d$ mixed state $\rho _{AC}$, the
R\'{e}nyi-$\alpha$ entanglement has an analytical expression
\beqa\label{q12}
E_\alpha  \left( {\rho _{AC} } \right) = f_\alpha  \left[ {C^2 \left( {\rho _{AC} } \right)} \right]
\eeqa
where the order $\alpha$ ranges in the region $[(\sqrt 7  - 1)/2,(\sqrt {13}  - 1)/2]$.

\emph{Proof}. Suppose that the optimal decomposition for $E_\alpha({\rho_{AC}})$ is $\{p_i, \ket{\psi_i}
_{AC}\}$, we have
\beqa\label{q13}
 E_\alpha \left( {\rho _{AC} } \right) &=& \sum\limits_i {p_i } E_\alpha  \left( {\left| {\psi ^i }
 \right\rangle _{AC} } \right) \nonumber\\
  &=& \sum\limits_i {p_i } E_\alpha  \left[ {C^2 \left( {\left| {\psi ^i } \right\rangle _{AC} } \right)}
  \right] \nonumber\\
  &\le& \sum\limits_i {q_i } E_\alpha  \left[ {C^2 \left( {\left| {\varphi ^i } \right\rangle _{AC} }
  \right)} \right] \nonumber\\
  &\le& E_\alpha  \left[ {\sum\limits_i {q_i } C^2 \left( {\left| {\varphi ^i } \right\rangle _{AC} }
  \right)} \right] \nonumber\\
  &=& E_\alpha  \left[ {C^2 \left( {\rho _{AC} } \right)} \right]
\eeqa
where in the third inequality we have assumed that $\left\{{q_i,\left|{\varphi ^i }\right\rangle_{AC}}
\right\}$ is the optimal decomposition for the squared concurrence $C^2 \left({\rho _{AC}}\right)$, and in
the fourth inequality we have used the property that $E_\alpha$ is a concave function of the squared
concurrence ${C^2 }$ for the order $\alpha\in[(\sqrt 7  - 1)/2, ( \sqrt {13}  - 1)/2]$ (lemma 3).

On the other hand, we can derive
\beqa\label{q14}
E_\alpha \left({\rho _{AC} } \right) &=& \sum\limits_i {p_i } E_\alpha \left( {\left| {\psi ^i }
\right\rangle _{AC} } \right) \nonumber\\
  &=& \sum\limits_i {p_i } E_\alpha  \left[ {C\left( {\left| {\psi ^i } \right\rangle _{AC} } \right)}
  \right] \nonumber\\
  &\ge& E_\alpha  \left[ {\sum\limits_i {p_i } C\left( {\left| {\psi ^i } \right\rangle _{AC} } \right)}
  \right] \nonumber\\
  &\ge& E_\alpha  \left[ {\sum\limits_k {r_k } C\left( {\left| {\phi ^k } \right\rangle _{AC} } \right)}
  \right] \nonumber\\
  &=& E_\alpha  \left[ {C\left( {\rho _{AC} } \right)} \right]
\eeqa
where the third inequality holds due to the convex property of $E_\alpha  \left( C \right)$
as a function of concurrence $C$ for $\alpha\geq (\sqrt7 - 1)/2$ \cite{kim10jpa,yxwang15arx}, and the
fourth inequality is satisfied due to $\{ r_k ,\ket{\phi^k}_{AC}\}$ being the optimal pure-state
decomposition for ${C({\rho_{AC}})}$.

Combining Eq. (\ref{q13}) with Eq. (\ref{q14}), we have $E_\alpha \left[ {C\left( {\rho _{AC} } \right)}
\right] \le E_\alpha  \left[ {\rho _{AC} } \right] \le E_\alpha  \left[ {C^2 \left( {\rho _{AC} }
\right)}\right]$. Due to $E_\alpha  \left[ {C\left( {\rho _{AC} } \right)} \right]$ and $E_\alpha
\left[ {C^2 \left( {\rho _{AC} } \right)} \right]$ being the same expression, we obtain the equality shown
in Eq. (\ref{q12}) and the proof is completed.

\emph{Corollary 1.} For the order $\alpha>(\sqrt{13}-1)/2\simeq 1.303$, the R\'{e}nyi-$\alpha$ entanglement
in bipartite $2\otimes d$ systems obeys the following relation
\begin{equation}\label{q15}
E_\alpha(\rho_{AC})\geq f_\alpha [ {C^2 ( {\rho _{AC}})}],
\end{equation}
which provides a nontrivial lower bound for the entanglement.

The proof of this corollary is straightforward according to Eq. (\ref{q14}).

\emph{Theorem 3}. For an arbitrary $N$-qubit mixed state $\rho _{A_1 A_{2 \ldots } A_N }$, there exist a set
of $k$-partite hierarchical monogamy relations
\beqa\label{q16}
E_\alpha ^2 \left( {\rho _{A_1 |A_{2 \ldots } A_N } } \right) \!\ge \!\sum\limits_{i = 2}^{k - 1}
{E_\alpha^2 \left({\rho_{A_1 A_i}} \right)} \! + \!E_\alpha^2\left({\rho _{A_1|A_{k \ldots }A_N}}\right)
\eeqa
where the number of parties is $k = \left\{ {3,4, \ldots ,N} \right\}$ and the order $\alpha$ in the region
[$(\sqrt 7  - 1)/2, ( \sqrt {13}  - 1)/2]$.

\emph{Proof}. We first consider a tripartite $2\otimes 2\otimes 2^{N-2}$ mixed state $\rho_{ABC}$, for
which we can derive
\begin{flalign}\label{q17}
 &E_\alpha ^2 \left( {\rho _{A|BC} } \right) - E_\alpha ^2 \left( {\rho _{AB} } \right) - E_\alpha ^2
 \left( {\rho _{AC} } \right) &\nonumber\\
  &= f_\alpha ^2 \left[ {C^2 \left( {\rho _{A|BC} } \right)} \right] - f_\alpha ^2 \left[ {C^2 \left( {\rho
  _{AB} } \right)} \right] - f_\alpha ^2 \left[ {C^2 \left( {\rho _{AC} } \right)} \right] &\nonumber\\
  &\ge f_\alpha ^2 \left[ {C^2 \left( {\rho _{AB} } \right) + C^2 \left( {\rho _{AC} } \right)} \right]
  &\nonumber\\
  &-f_\alpha ^2 \left[ {C^2 \left( {\rho _{AB} } \right)} \right]- f_\alpha ^2
 \left[ {C^2 \left( {\rho _{AC} } \right)} \right] &\nonumber\\
  &\ge 0&
\end{flalign}
where in the first equality we have used formula (\ref{q12}) in theorem 2, in the second inequality we have
utilized the monotonic property of  $f_\alpha ^2 \left( {C^2 } \right)$ and the monogamy relation of the
SC ${C^2 \left({\rho_{A|BC}}\right)\ge C^2\left({\rho_{AB}}\right)+C^2 \left({\rho _{AC}} \right)}$
\cite{osb06prl}, and in the third inequality we used the convex property of $f_\alpha ^2 ( {C^2 })$ in
lemma 2.

After further cutting the subsystem $C$ into a qubit $C_1$ and a qudit $C_2$ with $d=2^{N-3}$ and applying
Eq. (\ref{q17}) to the tripartite quantum state $\rho_{AC_1C_2}$, we can obtain
\begin{equation}
E_\alpha ^2 (\rho_{A|BC})-E_\alpha^2(\rho_{AB})-E_\alpha^2(\rho_{AC_1})-E_\alpha^2(\rho_{AC_2})\geq 0.
\end{equation}
By the successive cut for the last party and application of the tripartite monogamy inequality, we can
derive a set of hierarchical $k$-partite monogamy relations with $k = \left\{ {3,4, \ldots ,N} \right\}$
as shown in Eq. (\ref{q16}), and such that we prove theorem 3.

\section{Multipartite entanglement indicators based on the SR$\alpha$E and its applications}

According to the established monogamy relations based on the R$\alpha$E in Eqs. (8) and (16), we can construct
two kinds of multipartite entanglement indicators
\beqa\label{q19}
\tau _{\alpha(N)} ^{\left( 1 \right)}(\rho)  &=& \min \sum\limits_i {p_i } \tau_{\alpha(N)} \left(
{\left| {\psi _{A_1| A_2  \ldots A_n }^i } \right\rangle } \right)\\
\tau _{\alpha(K)} ^{\left( 2 \right)}(\rho)  &=& E_\alpha ^2 \left( {\rho _{A_1 |A_2  \ldots A_n } } \right)
- \sum\limits_{i = 2}^{k - 1} {E_\alpha ^2 \left( {\rho _{A_1 |A_i } } \right)} \nonumber\\
&-& E_\alpha ^2 \left( {\rho _{A_1 |A_k  \ldots A_n } } \right),
 \label{q20}
\eeqa
which can be utilized to detect the multipartite entanglement in an $N$-qubit state $\rho_{A_1A_2\dots A_n}$.
The first indicator $\tau _{\alpha(N)} ^{\left( 1 \right)}$ can detect the existence of multipartite
entanglement in an $N$-partite system, which comes from the convex roof of pure state indicator $\tau
_{\alpha(N)}  \left( {\ket{\psi} _{A_1 |A_{2 \cdots } A_n } } \right) = E_\alpha ^2 \left( {\left| \psi
\right\rangle _{A_1 |A_{2 \cdots } A_n } } \right) \!- \!\sum \limits_{i = 2}^{n} {E_\alpha ^2
\left( {\rho _{A_1 A_i } } \right)}$ with the minimum being taken over all possible pure state decompositions.
The second indicator $\tau _{\alpha(K)} ^{\left( 2 \right)}$ can detect the multipartite entanglement
in a $K$-partite system with $K\in\{3,4,\dots, N\}$, which is the residual entanglement in the hierarchical
monogamy inequality shown in Eq. (20). In the following, we give three examples as applications of the
above entanglement indicators.

\emph{Example 1}. We consider a three-qubit pure state $\left| {\psi\left( p \right)}\right\rangle=\sqrt p
\left| {GHZ_3} \right\rangle-\sqrt {1 - p} \left| W_3 \right\rangle$ which is the superposition of a $GHZ$
state and a $W$ state with $\ket{GHZ_3}=({\left|{000}\right\rangle+\left|{111}\right\rangle})/\sqrt{2}$
and $\ket{W_3}=(\ket{001}+\ket{010}+\ket{100})/\sqrt{3}$. The three-tangle $\tau$ is a tripartite
entanglement measure based on the monogamy relation of the SC and defined as
$\tau(\ket{\psi}_{ABC})=C_{A|BC}^2-C_{AB}^2-C_{AC}^2$ \cite{ckw00pra}. For the quantum state $\ket{\psi(p)}$,
its three-tangle is $\tau(\ket{\psi(p)})=p^2-8\sqrt{6}\sqrt{p(1-p)^3}/9$ which has two zero points at
$p_1=0$ and $p_2\simeq 0.627$ resulting in some flaw for the entanglement detection \cite{loh06prl,bxw14prl}.
In this case, we use the newly introduced multipartite entanglement indicator $\tau_{\alpha}^{(1)}$ shown
in Eq. (\ref{q19}). It is direct to calculate the value of $\tau_{\alpha}^{(1)}[\ket{\psi(p)}]$ since the
R$\alpha$E has an analytical formula for two-qubit quantum states and the convex roof extension is not
needed for the pure state case. In Fig.1, we plot the three-tangle and the indicator $\tau_{\alpha}^{(1)}$
for the order $\alpha=0.83, 1, 1.1$. As shown in the figure, the indicator $\tau_{\alpha}^{(1)}$ is always
positive for the different order $\alpha$ in contrast to the three-tangle $\tau_3$ having two zero points,
which detects all the genuine tripartite entanglement in all the region $p\in [0,1]$.

\begin{figure}[ptb]
\includegraphics[scale=0.7,angle=0]{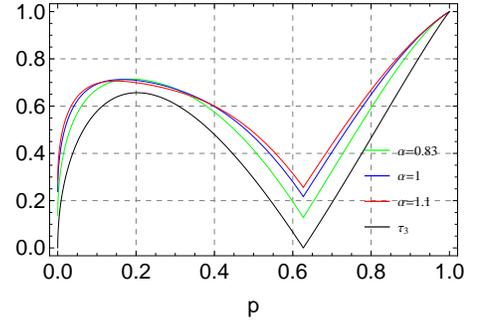}\caption{(color online). The indicator $
\tau _{\alpha (3)}^{\left( 1 \right)}$ for the superposition state $\left| {\psi\left( p \right)}\right
\rangle$ with $\alpha = 0.83$ (green line), $\alpha = 1$ (blue line), and $\alpha = 1.1$ (red line).
As an comparison, we also plot the three-tangle of $\left| {\psi\left( p \right)}\right\rangle$ with black line.}%
\label{fig1}
\end{figure}

\begin{figure}[ptb]
\includegraphics[scale=0.35,angle=0]{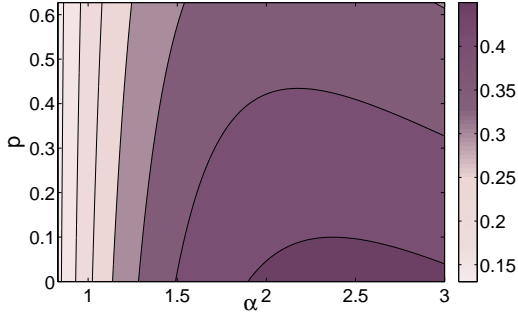}\caption{(color online). The indicator
$\tau _{\alpha (3)}^{\left( 1 \right)}$  with the order $\alpha \in (0.83,  3)$ detects the
existence of the genuine three-qubit entanglement in the mixed state $\rho_{ABC}$ when the
parameter $p\leq p_0$.}%
\label{fig2}
\end{figure}

\emph{Example 2}. Lohmayer \emph{et al} found that there exists a kind of three-qubit mixed states
\cite{loh06prl},
\begin{equation}\label{q21}
\rho_{ABC}  = p\left| {GHZ_3 } \right\rangle \left\langle {GHZ_3 } \right| + \left( {1 - p} \right)
\left| {W_3 } \right\rangle \left\langle {W_3 } \right|
\end{equation}
which is entangled but without two-qubit concurrence and three-tangle for the parameter $p\in[0.292,0.627]$.
Now we use the indicator $\tau_{\alpha}^{1}$ to detect the genuine three-qubit entanglement in the mixed
state. After some analysis, we obtain that the optimal pure state decomposition for $p \leq 0.627$ is
$\rho_{ABC}=(F/3)\sum\limits_{j = 0}^2 {\left| {\psi ^j \left( {p_0 } \right)} \right\rangle \left\langle
{\psi ^j \left( {p_0 } \right)} \right|}  + \left( {1 - F } \right)\left| {W_3 } \right\rangle \left\langle
{W_3 } \right|$ in which $F=p /p_0$ with $p_0=0.627$ and the component $\left| {\psi ^j \left( {p_0 } \right)}
\right\rangle  = \sqrt {p_0 } \left| {GHZ_3 } \right\rangle - e^{(2\pi i/3)j} \sqrt {1 - p_0 } \left| {W_3 }
\right\rangle$. Then we can derive $\tau _\alpha ^{\left( 1 \right)} \left( {\rho _{A|BC} } \right) = F \tau
_\alpha ^{\left( 1 \right)} \left( {\left| {\psi ^0 \left( {p_0 } \right)} \right\rangle } \right) +
\left( {1 - F } \right)\tau _\alpha ^{\left( 1 \right)} \left( {\left| {W_3 } \right\rangle } \right)$.
In Fig.\ref{fig2}, we plot the the indicator $\tau _\alpha ^{\left( 1 \right)} \left( {\rho _{ABC} } \right)$
as a function of parameters $p$ and $\alpha$. As shown in the figure, the values of this set of indicators are
always positive, which detect the existence of the genuine three-qubit entanglement in the mixed state. It is
noted that the case with the order $\alpha=1$ coincides with the result of the SEF-based indicator
\cite{bxw14prl} since the R$\alpha$E converge to the EOF in this case.

\emph{Example 3}. The three-tangle based on the monogamy relation of the SC cannot detect the tripartite
entanglement in the $W$ state. However, the SR$\alpha$E-based indication $\tau_{\alpha}^{(2)}$ can still
work in this case. We consider the $N$-qubit $W$ state in the form $\left| {W_N } \right\rangle =
1/\sqrt N  \left( {\left| {10 \cdots 0} \right\rangle  + \left| {01 \cdots 0} \right\rangle  +  \cdots
+ \left| {00 \cdots 1} \right\rangle } \right)$. When the quantum system is divided into $K$ parties with
$K\in[3,4,\dots N]$, there are a set of hierarchical monogamy relations. The corresponding indicator can be
written as $\tau _{\alpha \left( K \right)} ( {\left| {W_N } \right\rangle } )= E_\alpha ^2
( {C_{A_1 |A_{2 \cdots } A_N }^2 } ) - ( {k - 2} )E_\alpha ^2 ( {C_{A_1 A_2 }^2 } ) - E_\alpha ^2
( {C_{A_1 |A_{k \cdots } A_N }^2 } )$, where $C_{A_1 |A_{2 \cdots } A_N }^2  = 4(N - 1)/ N^2,
C_{A_1 A_2 }^2  = 4/N^2$, and $C_{A_1 |A_{k \cdots } A_N }^2 = 4(N - k + 1)/ N^2$. In Table I, we calculate
the indicator $\tau_{\alpha(K)}^{(2)}$ for a $7$-qubit $W$ state, where the party number $k$ ranges in
$[3,7]$ and the order $\alpha$ is chosen as 0.95, 1, 1.05, 1.1, and 1.15. The nonzero values of this
indicator reveal the multipartite entanglement in the $W$ state.

\begin{table}
\begin{ruledtabular}
\begin{tabular}{c|c|c|c|c|c}
&$\alpha=0.95 $ & $\alpha=1$ & $\alpha=1.05$ & $\alpha=1.10$
& $\alpha=1.15$ \\
\hline
$k=3$ & $0.0600$ & $0.0626$ & $0.0644$ & $0.0656$ & $0.0662$ \\
$k=4$ & $0.1136$ & $0.1178$ & $0.1205$ & $0.1219$ & $0.1225$ \\
$k=5$ & $0.1594$ & $0.1642$ & $0.1669$ & $0.1680$ & $0.1678$ \\
$k=6$ & $0.1954$ & $0.2000$ & $0.2021$ & $0.2023$ & $0.2010$ \\
$k=7$ & $0.2181$ & $0.2219$ & $0.2231$ & $0.2222$ & $0.2199$ \\
\end{tabular}
\end{ruledtabular}
\raggedright

\caption{The values of the indicator $\tau _{\alpha \left( K \right)} \left( {\left| {W_7 }
\right\rangle } \right)$ for the different party number $k$ and entanglement order $\alpha$.\label{tab1}}
\end{table}

\section{Discussion and conclusion}

We have considered the monogamy relations for the SR$\alpha$E in multiqubit systems. However, it is still an
open problem that whether this result can be extended to the multi-level systems. Ou pointed out that the
SC is not monogamous in a three-qutrit quantum state $\left| \psi  \right\rangle _{ABC} = \left(
{\left| {123} \right\rangle } \right. - \left| {132} \right\rangle  + \left| {231} \right\rangle -
\left| {213} \right\rangle \left. { + \left| {312} \right\rangle  - \left| {321} \right\rangle }
\right)/\sqrt{6}$ \cite{ou07pra}. When we use the monogamy relation of the SR$\alpha$E, it is found that
\beqa\label{q22}
E_\alpha ^2 ({\left| \psi  \right\rangle _{A|BC}}) - E_\alpha ^2 \left( {\rho _{AB} } \right)
- E_\alpha ^2 \left( {\rho _{AC} } \right) \simeq 0.51211,
\eeqa
which is monogamous for an arbitrary value of the order $\alpha$. Next, we consider a four-partite mixed state
$\rho _{A_1 A_2 A_3 A_4 }$ in $2 \otimes d_2  \otimes d_3  \otimes d_4$ systems. Suppose that
$C_{A_1 |A_2 A_3 A_4 }^2 = 0.7$, and $C_{A_1 A_2 }^2 = C_{A_1 A_3 }^2=C_{A_1 A_4 }^2=0.35$. In this case,
neither the SC or the SEF is monogamous and we have
\begin{eqnarray}\label{q23}
&&C^2(\rho_{A_1 |A_2 A_3 A_4 })-\sum_{i=2}^4 C^2(\rho_{A_1 A_i})=-0.35\nonumber\\
&&E_f^2 (\rho_{A_1 |A_2 A_3 A_4})-\sum_{i=2}^4 E_f^2(\rho_{A_1 A_i})=-0.037.
\end{eqnarray}
But the monogamy relation of the SR$\alpha$E still works for this mixed state and we can get
\beqa\label{q24}
E_\alpha ^2 ({\rho_{A_1|A_2 A_3 A_4}}) - \sum_{i = 2}^4 {E_\alpha ^2 ({\rho_{A_1 A_i}})} =0.052,
\eeqa
where the order $\alpha=1.2$ has been chosen.

Beside the monogamy relations we have established in terms of the SR$\alpha$E, the similar relations can
also be generalized to the $\mu$-th power of the R$\alpha$E. After some derivation, we can obtain the
following theorem and its proof can be seen from Appendix C.

\emph{Theorem 4}. For an arbitrary three-qubit mixed state ${\rho _{A_1 A_2 A_3 } }$, the $\mu$-th power
R\'{e}nyi-$\alpha$ entanglement obeys the monogamy relation
\beqa\label{q25}
E_\alpha^\mu  \left( {\rho _{A_1 |A_2 A_3 } } \right) \ge E_\alpha^\mu  \left({\rho _{A_1 A_2}}\right)
+ E_\alpha^\mu  \left( {\rho _{A_1 A_3 } } \right)
\eeqa
where the order $\alpha\ge(\sqrt{7}-1)/2\simeq 0.823$ and the power $\mu \geq 2$. Moreover, in $N$-qubit
systems, the following monogamy relation is also satisfied
\beqa\label{q26}
E_\alpha ^\mu  ( {\rho _{A_1 |A_{2 \ldots } A_n } } ) \ge \sum\limits_{i = 2}^{k - 1} {E_\alpha ^\mu
({\rho _{A_1 A_i }})}+E_{\alpha}^{\mu}(\rho_{A_1|A_k\dots A_n})
\eeqa
where the power $\mu\geq 2$ and the order $\alpha\in[\sqrt{7}-1)/2,(\sqrt{13}-1)/2]$.

We have thus proved the monogamy relation of the SR$\alpha$E in an arbitrary multi-qubit systems.
Our results provide a broad class of new monogamy inequalities which include the previous result in
terms of the SEF as a special case. Moreover, we have proved that the monogamy relations of the
SR$\alpha$E possess a hierarchical structure when the $N$-qubit system is divided into $k$ parties.
These new derived monogamy relations can be used to construct multipartite entanglement indicators in
$N$-qubit systems, which still work well even when the corresponding ones based on the SC and SEF lose
their efficacy. We also derived an analytical expression for the R$\alpha$E as a function of the SC in
bipartite $2\otimes d$ systems. Finally, we analyze the monogamy property of the $\mu$-th power of
R\'{e}nyi-$\alpha$ entanglement. It is still an open problem yet to be answered that whether there
exists a monogamy relation for the SR$\alpha$E in higher-dimensional systems .

\section*{Acknowledgment}

The authors would like to thank Jun-Sheng Feng, and Yong-Sheng Tang for helpful discussions.
This work was supported by NSF-China under Grant Nos. 11575051, 11374085, 11274010, and 11274124, the
Specialized Research Fund for the Doctoral Program of Higher Education (no. 20113401110002), and the 136 Foundation of Hefei Normal University under Grant
No.2014136KJB04. Y.-K. Bai was also supported by the Hebei NSF under Grant No. A2016205215.

\appendix

\section{Proof of lemma 2}

\begin{figure}
\includegraphics[scale=0.65,angle=0]{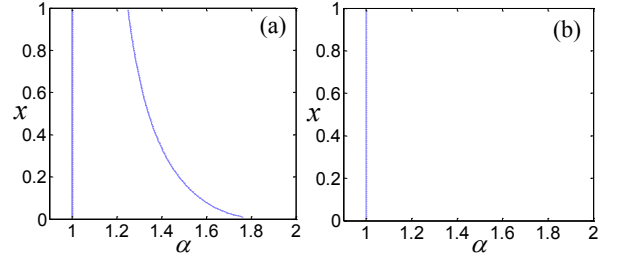}\caption{(color online). The plot of the dependence of
 $x$ with $\alpha$ which satisfies the equation (a)$\frac{{\partial h_\alpha  }}{{\partial x}} = 0$
 and (b)$\frac{{\partial h_\alpha  }}{{\partial \alpha }} = 0$ respectively.}%
\label{figa1}
\end{figure}

\begin{figure}
\includegraphics[scale=0.5,angle=0]{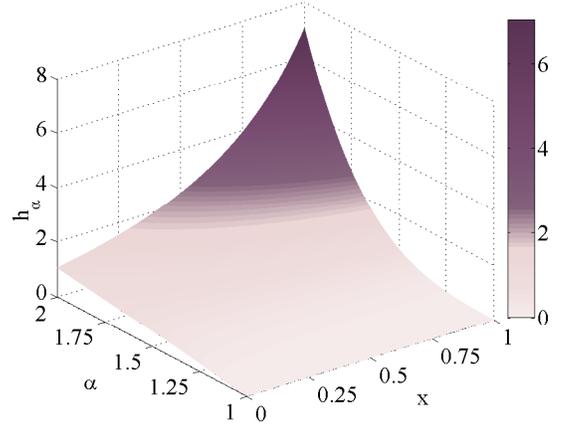}\caption{(color online). The function ${h_\alpha}$ is
plotted as a
function of $x$ and $\alpha$ for $0 \le x \le 1,\alpha  \ge 1$, which is positive, and as a
result, the SR$\alpha$E is a convex function of SC.}%
\label{figa2}
\end{figure}

This lemma holds if the second-order derivative $\partial ^2 E_\alpha^2 /\partial x^2>0$ for
$\alpha\geq (\sqrt{7}-1)/2$. We first consider the squared R\'{e}nyi-$\alpha$ entanglement for
$\alpha\geq 1$. In this case, we define a function
\beq\label{A1}
h_\alpha=\frac{{\partial ^2 \left[ {\left( {1 - \alpha } \right)^2 E_\alpha ^2 } \right]}}
{{\partial ^2 x^2 }}
\eeq
on the domain $\mathcal{D} = \left\{ {\left( {x,\alpha} \right)|0 \le x \le 1,\alpha\ge 1} \right\}$
with $x$ being the squared concurrence.
The nonnegativity of $h_\alpha$ can guarantee the nonnegative $\partial^2{E_\alpha^2}/\partial x^2$
since they are equivalent up to a positive constant. After some derivation, we have
\begin{align}\label{A2}
& h_\alpha   \!=\! \Lambda \left\{ {\alpha \left( {A^{\alpha  - 1}  \!-\! B^{\alpha  - 1} } \right)}
\right.^2\!+\! \left[ {\frac{{\left( {B^{\alpha-1}  \!-\! A^{\alpha-1}} \right)\left({A^\alpha\!+\!
B^\alpha  } \right)}}{{\sqrt {1 - x} }}} \right. &\nonumber\\
&\left. { \!+\! \left( {\alpha-1} \right)\left( {A^{\alpha-2}\!+\! B^{\alpha-2}} \right)\left({A^\alpha
\! +\! B^\alpha} \right)- \alpha \left( {A^{\alpha-1}  \!-\! B^{\alpha-1}}\right)^2} \right]
&\nonumber\\
&\left. {\times \ln \left[ {2^{ - \alpha } \left( {A^\alpha+ B^\alpha} \right)} \right]} \right\} &
\end{align}
where the parameter is $\Lambda=\alpha / 2( A^\alpha + B^\alpha )^2 (1 - x)(\ln 2)^2$ with
$A=1+\sqrt{1-x}$ and $B=1-\sqrt{1-x}$. For the proof of the nonnegativity of $h_\alpha$, it is
sufficient to analyze its maximal or minimal value on the domain $\mathcal{D}$. The critical points
of $h_\alpha$ satisfy the condition
\beqa\label{A3}
\nabla h_\alpha   = \left( {\frac{{\partial h_\alpha}}{{\partial x}},\frac{{\partial h_\alpha}}
{{\partial \alpha }}} \right) = 0.
\eeqa
In Fig.\ref{figa1} (a) and (b), we plot the solutions to equations $\partial h_\alpha/\partial x=0$
and $\partial h_\alpha/\partial \alpha= 0$, respectively. As shown in the figure, the common solution
is $\alpha=1$ which is on the boundary of the domain $\mathcal{D}$. Therefore, the maximal or minimal
value of $h_\alpha$ can arise only on the boundary of domain $\mathcal{D}$. Next, we consider the other
two boundaries $x=0$ and $x=1$ on the domain $\mathcal{D}$ of $h_\alpha$. When $x=0$, we have
\beqa\label{A4}
\mathop {\lim }\limits_{x \to 0} h_\alpha= \frac{{{\alpha}^2 }}{{8{\left( {\ln 2} \right)^2 }}}
\eeqa
which is always positive in the region $\alpha \in (1, + \infty )$. Similarly, when $x=1$, we can derive
\beqa\label{A5}
\mathop {\lim }\limits_{x \to 1} h_\alpha= \frac{{(1 - \alpha )^2 \alpha \left[ {3\alpha +
2\left( {\alpha ^2  + \alpha  - 3} \right)\ln 2} \right]}}{{6{\left( {\ln 2} \right)^2 }}}
\eeqa
which is monotonically increasing and positive in the region $\alpha \in (1, + \infty )$.
Notice that the critical points arise only on the boundary of domain $\mathcal{D}$, we obtain that
the function $h_\alpha$ is nonnegative in the whole range ${0 \le x \le 1,\alpha  \ge 1}$ (the
equality holds only at the boundary $\alpha=1$). In Fig. \ref{figa2}, we plot $h_\alpha$ as a function
of $x$ and $\alpha$, which illustrates our result. According to the equivalent relation in Eq.
(\ref{A1}), we have $\partial ^2 E_\alpha^2 /\partial x^2>0$ for $\alpha>1$. When $\alpha=1$,
$E^2_\alpha$ converges to SEF and its second-order derivative is positive \cite{bxw14prl}. Therefore,
the second-order derivative of $E^2_\alpha$ is positive for $\alpha\geq 1$.

\begin{figure}[b]
\includegraphics[scale=0.65,angle=0]{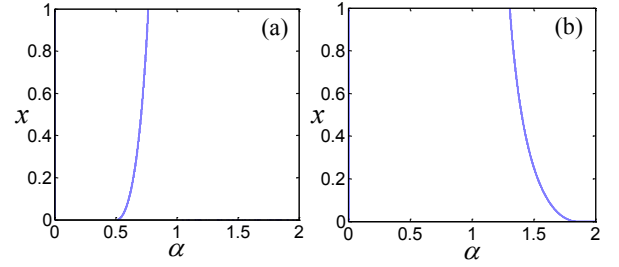}\caption{(color online). The plot of the dependence of $x$
with $\alpha$ using the equation (a)$\frac{{\partial^2 E_\alpha ^2 }}{{\partial x^2 }} = 0$,
(b)$\frac{{\partial^2 E_\alpha  }}{{\partial x^2 }} = 0$.}%
\label{figa3}
\end{figure}

We further analyze the nonnegative region for the second-order derivative $\partial ^2 E_\alpha ^2
/\partial x^2$ when $\alpha$ ranges in $(0,1)$. It is found that, under the condition
$\partial ^2 E_\alpha ^2 /\partial x^2=0$, the critical value of $x$ increases monotonically along
with the parameter $\alpha$. In Fig.\ref{figa3}(a), we plot the solution $(x,\alpha)$ to this critical
condition, where for each fixed $x$ there exists a value of $\alpha$ such that the second-order
derivative of $E_\alpha ^2$ is zero. Due to $x$ varying monotonically with $\alpha$, we only need
consider the condition $\partial ^2 E_\alpha  ^2 /\partial x^2 = 0$ in the limit $x \to 1$. In this
case, we have the derivative
\beqa\label{A6}
\mathop {\lim }\limits_{x \to 1} \frac{{\partial ^2 E_\alpha^2 }}{{\partial x^2 }}=\frac{{\alpha
\left[{3\alpha+ 2\left({\alpha^2+\alpha-3}\right)\ln 2}\right]}}{{6{\left({\ln 2}\right)^2}}}=0,
\eeqa
which gives the critical point $\alpha_{c_1}=[-(2\ln2 +3)+\sqrt{(2\ln2 +3)^2+48(\ln2)^2}]/4\ln2\simeq
0.764$. When $\alpha \ge \alpha _{c_1}$, the second-order $\partial ^2 E_\alpha^2 /\partial x^2$
is always positive. Notice that the analytical formula $E_\alpha(C^2)$ in Eq.(\ref{q5}) is established
only for $\alpha \geq (\sqrt7-1)/2=\alpha_c$, we have $\partial ^2 E_\alpha^2 /\partial x^2>0$ for
$\alpha\in[\alpha_c,1)$.

Combining the two positive regions $[\alpha_c,1)$ and $[1,+\infty)$, we obtain the derivative
$\partial ^2 E_\alpha^2 /\partial x^2>0$ for $\alpha\geq (\sqrt{7}-1)/2$, which completes the proof
of lemma 2.

\section{Proof of lemma 3}

The R$\alpha$E $E_\alpha$ is monotonically increasing if the first-order derivative $\partial E_\alpha
/\partial x > 0$ with $x$ being the squared concurrence. After some calculation, we have
\beqa\label{B1}
\frac{{\partial E_\alpha  }}{{\partial x}}=\frac{{\alpha\left({B^{\alpha - 1}-A^{\alpha-1}} \right)}}
{{2\left( {1 - \alpha} \right)\left( {A^\alpha  + B^\alpha } \right)\sqrt{1-x} \ln 2}}
\eeqa
where the parameters are $A=1+\sqrt{1-x}$ and $B=1-\sqrt{1-x}$. It is easy to verify the derivative
is nonnegative in the regions $\alpha  \in \left[ {0,1} \right)$ and $\alpha\in\left( {1, +\infty}\right)$.
Combining the two regions with the case $\alpha=1$ which was proved in Ref. \cite{bxw14pra}, we obtain
that the first-order derivative of ${E_\alpha }$ is always nonnegative for $\alpha\geq 0$ and the equality
holds only at the boundary of $x$. Therefore, the R$\alpha$E is monotonically increasing.

Furthermore, the concave property of ${E_\alpha }$ as a function of ${C^2 }$ holds if the second-order
derivative $\partial^2 E_\alpha/\partial x^2< 0$ with $x$ being the squared concurrence. In order to
determinate the region of $\alpha$, we analyze the condition $\partial^2 E_\alpha/\partial x^2=0$. It is
found that the value of $x$ decreases monotonically along with the increase of $\alpha$. In Fig.
\ref{figa3}(b), we plot the relation between $x$ and $\alpha$ under this condition. As shown in the
figure, the critical point corresponds to the limit
\beqa\label{B2}
\mathop {\lim }\limits_{x \to 1} \frac{{\partial^2 E_\alpha  }}{{\partial x^2 }} = \frac{{\alpha
\left( {\alpha ^2  + \alpha  - 3} \right)}}{{6\ln 2}} = 0.
\eeqa
After some calculation, we can obtain that the critical point is $\alpha_{c_2}=(\sqrt {13}-1)/2\simeq
1.303$. Therefore, the second-order derivative is negative when $\alpha<\alpha_{c_2}$. It is noted that
the analytical formula $E_\alpha(C^2)$ hold for $\alpha\geq (\sqrt{7}-1)/2$. Thus we can get $\partial^2
E_\alpha/ \partial x^2\leq 0$ for $\alpha\in[(\sqrt{7}-1)/2,(\sqrt {13}-1)/2]$ (the equality holds only
at the right boundary), which results in the concave property of $E_\alpha(C^2)$. The proof of lemma 3
is completed.

\section{Proof of theorem 4}

According to theorem 1, we have
\beqa\label{C1}
E_\alpha^2({\rho_{A_1|A_2A_3}})\geq E_\alpha^2({\rho_{A_1A_2}})+E_\alpha^2 ({\rho_{A_1A_3}})
\eeqa
for $\alpha \geq(\sqrt{7}-1)/2$. Without loss of generality, we assume the two-qubit entanglement
$E_\alpha({\rho_{A_1A_2}})\geq E_\alpha ({\rho_{A_1A_3}})$. Then we can get
\beqa\label{C2}
E_\alpha^\mu  \left({\rho _{A_1|A_2A_3}}\right) &\ge&\left( {E_\alpha^2 \left({\rho_{A_1A_2}}\right) +
E_\alpha^2 \left( {\rho _{A_1 A_3 } } \right)} \right)^{\frac{\mu }{2}}  \notag\\
&=& E_\alpha^\mu \left({\rho_{A_1A_2}} \right)\left({1+\frac{{E_\alpha^2 \left({\rho_{A_1A_3}}
\right)}}{{E_\alpha^2 \left( {\rho _{A_1 A_2 } } \right)}}} \right)^{\frac{\mu }{2}} \notag\\
&\ge& E_\alpha^\mu  \left( {\rho _{A_1 A_2 } } \right)\left( {1 + \left( {\frac{{E_\alpha^2
\left( {\rho _{A_1 A_3 } } \right)}}{{E_\alpha^2 \left( {\rho _{A_1 A_2 } } \right)}}} \right)
^{\frac{\mu }{2}} } \right) \notag\\
&=& E_\alpha^\mu\left({\rho_{A_1A_2}}\right)+E_\alpha^\mu\left({\rho_{A_1A_3 }}\right)
\eeqa
where in the third inequality we have used the property $\left( {1 + x} \right)^t  \ge 1 + x^t$,
for $x \le 1,t \ge 1$.

For the monogamy relation of $\mu$-th power in $N$-qubit systems, we first consider a tripartite mixed state
in $2 \otimes 2 \otimes 2^{N - 2}$ systems. According to theorem 3, the tripartite monogamy relation for the
squared R$\alpha$E is satisfied. Then, using the same technique in Eq. (\ref{C1}), we can obtain the $\mu$-th
power monogamy inequality. Furthermore, by the successive cut for the last party and application of the
tripartite monogamy relation, we can derive the $k$-partite inequality as shown in Eq. (\ref{q26}) of the
main text.

\end{document}